\documentclass[a4paper,11pt]{article}
\pdfoutput=1 

\usepackage{jheppub} 

\usepackage[T1]{fontenc} 
\usepackage{slashed}

\title{\boldmath Constraining CP-phases in SUSY: an interplay of muon/electron $g-2$ and electron EDM}

\author[a,b]{Song Li,}
\author[a,b]{Yang Xiao,}
\author[a,b]{Jin Min Yang}

\affiliation[a]{CAS Key Laboratory of Theoretical Physics, Institute of Theoretical Physics, Chinese Academy of Sciences, Beijing 100190, China}
\affiliation[b]{School of Physics Sciences, University of Chinese Academy of Sciences,  Beijing 100049, China}

\emailAdd{lisong@itp.ac.cn}
\emailAdd{xiaoyang@itp.ac.cn}
\emailAdd{jmyang@itp.ac.cn}

\abstract{
  The minimal supersymmetric standard model (MSSM) with complex parameters can contribute sizably to muon/electron anomalous magnetic dipole momemnt ($g-2$) and  electric dipole moment (EDM).  The electron $g-2$ interplays with electron EDM; the muon $g-2$ can also interplay with electron EDM assuming the universality between smuon and selectron masses, either of which can constrain the relevant CP-phases in the MSSM.  
  In this work, we first use such an interplay to derive an approximate analytical upper limit on the relevant CP-phase.
  Then we extensively scan the parameter space to obtain more accurate upper limits.
  We obtain the following observations: 
  (i) The muon $g-2$ in the $2\sigma$ range combined with the electron EDM upper limit (assuming the universality between smuon and selectron masses) typically constrains the relevant CP-phase under $1.9\times 10^{-5} (\text{rad})$;
  (ii) The electron $g-2$ in the $2\sigma$ range (Berkeley) interplays with the electron EDM upper limit (without assuming the universality between smuon and selectron masses) constrains the relevant CP-phase under $3.9\times 10^{-6}(\text{rad})$ (also requiring muon $g-2$ in the allowed $2\sigma$ range). We also find some special cancellations in the parameter space which can relax the constraints by a couple of orders. 
  Such stringent limits on CP-phases may pose a challenge for model building of SUSY, i.e., how to naturally suppress these phases. 
}

\begin{document} 
\maketitle
\flushbottom

   \section{Introduction}
   \label{sec:intro}
   The latest measurement of muon $g-2$ from Fermilab~\cite{PhysRevLett.126.141801} gave a value that exceeds the Standard Model (SM) prediction (see~\cite{Aoyama:2020ynm} and refs therein) by $4.2\sigma$ after combining with the BNL result~\cite{Muong-2:2006rrc}:
   \begin{equation}\label{eq:muonGM2}
       \Delta a_{\mu}^{\rm{Exp-SM}} =a_\mu^{\rm Exp}-a_\mu^{\rm SM}= (2.51\pm 0.59) \times 10^{- 9}.
   \end{equation}
   For electron $g-2$, if we deduce it from the fine structure constant $\alpha_{\rm em}$ measured by Berkeley using $^{133}{\rm Cs}$ atoms~\cite{Parker:2018vye},  we will find that the experimental value~\cite{Hanneke:2008tm} is $2.4\sigma$ below the SM prediction~\cite{Aoyama:2019ryr}:
   \begin{equation} \label{eq:electronGM2}
      \Delta a_e^{\rm{Exp-SM}}=a_e^{\rm Exp}-a_e^{\rm SM}(\text{Cs})=(-8.8\pm 3.6)\times 10^{- 13}.
   \end{equation}
   However, the $\alpha_{\rm em}$ measurement by Laboratoire Kastler Brossel (LKB)~\cite{Morel:2020dww,Aoyama:2012wj} using $^{87}{\rm Rb}$ leads to a prediction value of electron $g-2$ that agrees well with the experimental value. So far the reason that caused the deviation between the two experiments is unkown.
   
   In the minimal supersymmetric standard model (MSSM), a simultaneous explanation of the muon $g-2$ and the LKB electron $g-2$ is not hard (since no SUSY effects are needed for the LKB electron $g-2$), while a simultaneous explanation of the muon $g-2$ and the Berkeley electron $g-2$ is rather challenging within the MSSM since this requires a positive contribution to the muon $g-2$ and a negative contribution to the electron $g-2$ ~\cite{Endo:2019bcj,Dutta:2018fge,Badziak:2019gaf,Li:2021koa,Crivellin:2018qmi} (in this work we restrain ourselves in the minimal framework of supersymmetry (SUSY),
   whereas in some extensions of the MSSM like the B-L SSM and an extended next-to-MSSM, or in some falvor models or 2HDM+S, a joint explanation is not so hard~\cite{Yang:2020bmh,Cao:2021lmj,Calibbi:2020emz,Keus:2017ioh}). 
 
   The current upper limit of the electron EDM is~\cite{ACME:2018yjb}
   \begin{equation}\label{eq:electronEDM}
       |d_e|<1.1\times 10^{-29}{\rm e~cm}~(\text{90\% C.L.}).
   \end{equation}
   This experimental upper limit strictly constrains the CP-phases of the new physics. In the MSSM, the CP-phases should be present since they are not forbidden by any fundamental symmetry, just as the CP-phases in the SM.  Although such CP-phases cannot explain the baryon asymmetry of the universe through the electroweak baryogenesis mechanism, phenomenologically we need to know the limits on these CP-phases from current experiments, and these limits may provide us some enlightenment for the UV completion of the MSSM. Note that in the MSSM the $g-2$ and EDM receive contributions from  the same loops. Therefore, the muon/electron $g-2$ and the electron EDM limit can be jointly used to constrain the CP-phasese in the MSSM. In ~\cite{Han:2021ify}, assumming the universality for the first and second generation sleptons, the muon $g-2$ and the electron EDM limit are found to set an upper bound of $\sim 10^{-5}$ on the relevant CP-phasese in the MSSM.  In this work, we will perform a more comprehensive research, considering 
  the following two situations within the MSSM:
   \begin{enumerate}
       \item Adopting the electron $g-2$ result of LKB, we will take the popular assumption that SUSY is universal for the first and second generation sleptons. In this case, the muon $g-2$ can be easily explained in the MSSM ~\cite{Martin:2001st}, which, combined with the electron electric dipole moment (EDM) limits, can set constraints on the CP-phases of the relevant parameters. This scenario was considered in ~\cite{Han:2021ify}. But we will perform an intensice scan and our results typically agree with ~\cite{Han:2021ify} (i.e. obtain an upper bound of $\sim 10^{-5}$ on the relevant CP-phasese in the MSSM) except that we find in some special cancellation cases the upper bound can be weakened by a couple of orders;  
       \item Adopting the electron $g-2$ result of Berkeley, we will assume that SUSY is not universal for the first and second generation sleptons. In this case, the MSSM parameter space for a joint explanation of the muon $g-2$ and the Berkeley electron $g-2$ ~\cite{Badziak:2019gaf,Li:2021koa}, combined with the electron EDM limits, can set rather strong constraints on the CP-phases of the relevant parameters (the upper bound is found to be typically $\sim 10^{-6}$).  
   \end{enumerate}
   Note that in  ~\cite{Kraml:2007pr} various SUSY explanations for the tiny electron EDM were listed, e.g., 
   heavy sparticles~\cite{Nath:1991dn,Kizukuri:1991mb,Kizukuri:1992nj,Cohen:1996vb,Falk:1995fk}, 
   accidental cancellations~\cite{Falk:1996ni,Falk:1998pu,Ibrahim:1997gj,Ibrahim:1997nc,Ibrahim:1998je,Brhlik:1998zn,Brhlik:1999ub,Bartl:1999bc,Pokorski:1999hz,Arnowitt:2001pm,Dhuria:2013syh}, 
   flavour off-diagonal CP-phases~\cite{Abel:2001vy} and  
   lepton flavour violation~\cite{Bartl:2003ju} (for some more recent attempts, see, e.g.,  ~\cite{Diaz-Cruz:2005uyk,Paradisi:2009ey,Moroi:2011fi,Ibe:2021cvf}).
  In this work, we do not discuss which explanation is most suitable. Instead, we only study the upper limits on the  CP-phases from the phenomenological view using the experimental results of muon/electron $g-2$ and the electron EDM limit. We do not consider the dark matter relic density and the collider constraints since they have little connection with the CP-phases.
   
   This work is organized as follows. In Sec.~\ref{sec2}, we describe the MSSM contributions to the muon/electron $g-2$ and electron EDM.  In Sec.~\ref{sec3}, we investigate the upper limits on the CP-phases in the MSSM. Finally, we conclude in Sec.~\ref{sec:conclusions}.

   \section{Muon/electron \texorpdfstring{$g-2$}{g-2} and EDM in the MSSM}\label{sec2}
   \subsection{Muon/electron \texorpdfstring{$g-2$}{g-2} in the MSSM}
   In the MSSM, the contribution of the supersymmetric particles to the lepton $g-2$ consists of two parts~\cite{Martin:2001st}:
   \begin{align}\label{eq:leptonMDM1}
	\delta a_{\ell}^{\chi^0} & =  \frac{m_{\ell}}{16\pi^2}
    \sum_{i,m}\left\{ -\frac{m_{\ell}}{ 12 m^2_{\tilde\ell_m}}
    (|n^L_{im}|^2+ |n^R_{im}|^2)F^N_1(x_{im}) 
    +\frac{m_{\chi^0_i}}{3 m^2_{\tilde\ell_m}}
    {\rm Re}[n^L_{im}n^R_{im}] F^N_2(x_{im})\right\}, \\
    \label{eq:leptonMDM2}
    \delta a_{\ell}^{\chi^\pm} & =  \frac{m_{\ell}}{16\pi^2}\sum_k
    \left\{ \frac{m_\ell}{ 12 m^2_{\tilde\nu_\ell}}
    (|c^L_k|^2+ |c^R_k|^2)F^C_1(x_k)
    +\frac{2m_{\chi^\pm_k}}{3m^2_{\tilde\nu_\ell}} {\rm Re}[ c^L_kc^R_k] F^C_2(x_k)\right\},
   \end{align}
   where $i= 1,2,3,4$,  $m= 1,2$ and $k= 1,2$ label respectively the neutralinos, sleptons and charginos in mass eigenstates, and $x_{im}=m^2_{\chi^0_i}/m^2_{\tilde \ell_m}$, $x_k=m^2_{\chi^{\pm}_k}/m^2_{\tilde \nu_\ell}$,
   \begin{align}
      \label{eq:nRim}
      n^R_{im} &= \sqrt2 g_1 N_{i1}X_{m2}+y_\ell N_{i3}X_{m1},\\
      \label{eq:nLim}
      n^L_{im} &= \frac{1}{\sqrt2}(g_2 N_{i2}+g_1 N_{i1})X^*_{m1}-y_\ell N_{i3}X^*_{m2},\\
      c^R_k &= y_\ell U_{k2},\\
      c^L_k &= -g_2 V_{k1},
   \end{align}
   with $y_\ell=g_2 m_\ell/(\sqrt2 m_W\cos\beta)$ being the lepton Yukawa coupling, $N$ and $(U,V)$ and $X$ being the mixing matrices for the neutralinos, charginos and sleptons, respectively. These matrices satisfy
   \begin{align}
       N^*M_{\chi^0}N^\dagger &= {\rm diag}(m_{\chi^0_1},m_{\chi^0_2},m_{\chi^0_3},m_{\chi^0_4}),\\
       U^*M_{\chi^\pm} V^\dagger &= {\rm diag}(m_{\chi^\pm_1},m_{\chi^\pm_2}), \\
       X M^2_{\tilde\ell} X^\dagger &= {\rm diag}(m^2_{\tilde\ell_1}, m^2_{\tilde\ell_2}).
   \end{align}
   The chargino mass matrix $M_{\chi^\pm}$, the neutralino mass matrix $M_{\chi^0}$, the slepton mass matrix $M^2_{\tilde\ell}$, and the definitions of $F_{1,2}^{N,C}$ in Eq.~\eqref{eq:leptonMDM1} and Eq.~\eqref{eq:leptonMDM2} can be found in ~\cite{Martin:2001st}.
   
   Furthermore, we can calculate $\delta a_{\ell}^{\rm SUSY}$ with high-loop corrections
   \begin{equation}\label{eq:MDM2loop}
      \delta a_\ell^{\rm SUSY}=\left(1-\frac{4\alpha}{\pi}\ln\frac{M_{\rm SUSY}}{m_\ell}\right)\left(\frac{1}{1+\Delta_\ell}\right)\delta a_\ell^{\rm SUSY,\,1L}.
   \end{equation}
    The detailed information of these corrections and the definition of $\Delta_l$ can be found in ~\cite{Degrassi:1998es,Marchetti:2008hw,Carena:1999py}. For the calculation of the EDM in the following, we will also consider similar high-loop corrections. 
    On the other hand, the definition of $\Delta_l$ involves a loop function containing false singularities. We use the technique in ~\cite{Li:2021koa} to avoid these false singularities. 
   
   \subsection{Electron EDM in the MSSM}
   In the MSSM, the electron EDM comes from the same loops as in $g-2$, which also consists of two parts~\cite{Ibrahim:2007fb,Cheung:2009fc}
   \begin{align}\label{eq:electronEDM1}
	  \frac{d_{e}^{\chi^0}}{\rm e} &= - \frac{1}{16\pi^2} \sum_{i,m} \frac{m_{\chi^0_i}}{6 m^2_{\tilde e_m}} {\rm Im}[n^L_{im}n^R_{im}] F^N_2(x_{im}), \\
      \label{eq:electronEDM2}
      \frac{d_{e}^{\chi^\pm}}{\rm e} &= - \frac{1}{16\pi^2} \sum_{k} \frac{m_{\chi^\pm_k}}{3m^2_{\tilde\nu_e}} {\rm Im}[ c^L_kc^R_k] F^C_2(x_k),
   \end{align}
   with ${\rm e}$ being the electric charge of the positron. We need to clarify the CP-phases involved in the electron EDM. The neutralino mass matrix and its Takagi decomposition is
   \begin{align}
       & M_{\chi^0} = \begin{pmatrix}
          M_1 & 0 & -m_Z c_\beta s_W & m_Z s_\beta s_W\\
          0 & M_2 & m_Z c_\beta c_W & -m_Z s_\beta c_W \\
          -m_Z c_\beta s_W & m_Z c_\beta c_W & 0 & -\mu\\
          m_Z s_\beta s_W & -m_Z s_\beta c_W & -\mu & 0
       \end{pmatrix} ,\\
       & N^*M_{\chi^0}N^\dagger = {\rm diag}(m_{\chi^0_1},m_{\chi^0_2},m_{\chi^0_3},m_{\chi^0_4}),
   \end{align}
   where $c_\beta=\cos\beta$, $s_\beta=\sin\beta$, $c_W=\cos\theta_W$ and $s_W=\sin\theta_W$ with $\theta_W$ being the weak mixing angle. We assume $\mu=|\mu|e^{i\varphi_\mu}$, $D={\rm diag}(e^{-i\varphi_\mu/2},e^{-i\varphi_\mu/2},e^{i\varphi_\mu/2},e^{i\varphi_\mu/2})$, and
   \begin{equation}
       M_{\chi^0}' = \begin{pmatrix}
          M_1e^{i\varphi_\mu} & 0 & -m_Z c_\beta s_W & m_Z s_\beta s_W\\
          0 & M_2e^{i\varphi_\mu} & m_Z c_\beta c_W & -m_Z s_\beta c_W \\
          -m_Z c_\beta s_W & m_Z c_\beta c_W & 0 & -|\mu|\\
          m_Z s_\beta s_W & -m_Z s_\beta c_W & -|\mu| & 0
       \end{pmatrix}.
   \end{equation}
   Hence we have $DM_{\chi^0}'D^{\rm T}=M_{\chi^0}$. Defining $\tilde N^*=N^*D$, we obtain
   \begin{equation}
       \tilde N^* M_{\chi^0}'\tilde N^\dagger={\rm diag}(m_{\chi^0_1},m_{\chi^0_2},m_{\chi^0_3},m_{\chi^0_4}).
   \end{equation}
   $M_{\chi^0}'$ (and thus $\tilde N$) only contains two phases, ${\rm Arg} (\mu M_1)$ and ${\rm Arg} (\mu M_2)$. And we know 
   \begin{equation}
       N=\tilde ND=\tilde N \begin{pmatrix}
          e^{-i\varphi_\mu/2} &   &   &  \\
            & e^{-i\varphi_\mu/2} &   &   \\
            &   & e^{i\varphi_\mu/2} &  \\
            &   &   & e^{i\varphi_\mu/2}
       \end{pmatrix}.
   \end{equation}
   Similarly, for the slepton mass matrix, we can obtain the matrix $\tilde X$ containing only the phase ${\rm Arg}(\mu A_e)$, and 
   \begin{equation}
       X=\tilde X \begin{pmatrix}
          e^{-i\varphi_\mu/2} &  \\
            & e^{i\varphi_\mu/2} 
       \end{pmatrix}.
   \end{equation}%
   Now from Eq.\eqref{eq:nRim} and Eq.\eqref{eq:nLim} we know that the phases which $n^R_{im}$ and $n^L_{im}$ depend on are ${\rm Arg} (\mu M_1)$, ${\rm Arg} (\mu M_2) $ and ${\rm Arg}(\mu A_e)$, hence $d^{\chi^0}_e/{\rm e}$ also depends on these three phase combinations~\cite{Ibrahim:2007fb,Cheung:2009fc}. However, the influence of ${\rm Arg}(\mu A_e)$ in the electron EDM will be less significant than the other two phase combinations due to the small electron mass. After a similar derivation, we will find that $d^{\chi^\pm}_e/{\rm e}$ depends on the phase ${\rm Arg} (\mu M_2)$~\cite{Ibrahim:2007fb,Cheung:2009fc}.

   \section{Upper limits on CP-phases in the MSSM} \label{sec3}
   Comparing the formulas of $g-2$ in Eq.\eqref{eq:leptonMDM1} and Eq.\eqref{eq:leptonMDM2} and the formulas of the electron EDM in Eq.\eqref{eq:electronEDM1} and Eq.\eqref{eq:electronEDM2}, we can find
   \begin{align}
       \frac{\delta a_\ell}{m_\ell} &= D_\ell+B_\ell\cos\varphi,\\
       \frac{d_e}{\rm e} &= -\frac12 B_e\sin\varphi,
   \end{align}
   where $D_\ell$ and $B_\ell$ have the same order of magnitude, and $\varphi$ is the dominant CP phase in $d_e/{\rm e}$. In this work, we take the CP phase in $(-\pi/2,\pi/2)$. Because of the term $D_\ell$, the value range of $\delta a_\ell$ is not greatly affected by the phase. Especially when the phase is small, $\delta a_\ell/m_\ell\approx D_\ell+B_\ell\approx 2B_\ell$, $d_e/{\rm e}\approx -B_e\varphi/2$. Therefore, through $\delta a_\ell$, one can constrain the range of $D_\ell$ and $B_\ell$. One can also get an upper limit on the phase according to the experimental value of $d_e$ if $B_e$ has a proportional relationship with one of $B_\ell(\ell=e,\,\mu)$. Assuming the lepton flavor universality in the MSSM, we then have ~\cite{book:Jegerlehner:2017}
   \begin{equation}\label{eq:delta-alepton}
      \delta a_{\ell}\propto \left(\frac{m_{\ell}}{M_{\rm SUSY}}\right)^2.
   \end{equation}
   Therefore,
   \begin{align}
       \delta a_\ell &\sim\left(\frac{m_\ell}{m_e}\right)^2\delta a_e\sim2\frac{m^2_\ell}{m_e}B_e,\\
       |d_e| &\sim \frac{1}{2}{\rm e}|B_e||\varphi|\lesssim d_e^{\rm lim},
   \end{align}
   where $d_e^{\rm lim}$ represents the experimental upper limit on the electron EDM.
   Combining these two formulas, we obtain
   \begin{equation}\label{eq:phaselimit}
       |\varphi|\lesssim\frac{4m_\ell^2}{m_e|\delta a_\ell|} \frac{d_e^{\rm lim}}{\rm e}.
   \end{equation}
    This result allows us to estimate the upper limits on the CP-phase from the values of  $d_e^{\rm lim}$ and $\delta a_\ell$ ($\ell=e,\mu$).

   \subsection{CP-phase upper limits with flavor universality}
   We calculate the phase limits from the interplay of $\delta a_\mu$ and the upper limit of the electron EDM,  assuming the lepton flavor universality. In this scenario we take the LKB value for the electron $g-2$ which agrees with the SM prediction ~\cite{Morel:2020dww,Aoyama:2012wj}. Since no SUSY effects are needed, the electron $g-2$ $2\sigma$-constraint can be trivially satisfied in this scenario.  
   Using Eq.\eqref{eq:phaselimit} to estimate the limit of the corresponding phase, we obtain
   \begin{equation}
       |\varphi|\lesssim\frac{4m_\mu^2}{m_e\delta a_\mu} \frac{d_e^{\rm lim}}{\rm e}\sim 2\times10^{-5}( \text{rad}),
   \end{equation}
   where we take the central value in Eq.\eqref{eq:muonGM2} for $\delta a_\mu$ and  the upper bound in Eq.\eqref{eq:electronEDM} for $d_e^{\rm lim}$. 
   From the following results of a parameter scan, we will find that this estimate is quite accurate for the dominant phase. 
   
   Now we perform a scan over the parameter space. In order to highlight the dominant phase effects, we need to decouple the unnecessary phases. For the situation where ${\rm Arg}(\mu M_2)$ dominates, we choose the following parameter scenario~\cite{Han:2021ify}:
   \begin{align*}
       &100~\text{GeV}<(|\mu|,|M_2|,M_{E1}=M_{E2},M_{L1}=M_{L2})<2~\text{TeV},\\
       &|M_1|=5~\text{TeV},\\
       &2<\tan\beta <50,\\
       &\text{Arg}(\mu M_1)=\text{Arg}(\mu A_e)=0,\\
       &-\frac{\pi}{2}<\text{Arg}(\mu M_2)<\frac{\pi}{2}, \\
       &A\text{ parameters} = 0.
   \end{align*}
   Here we set $|M_1|=5~{\rm TeV}$ to decouple $M_1$ from $\delta a_\mu$ and $d_e$. In this case, the SUSY contributions to the muon $g-2$ and the electron EDM mainly come from the wino-higgsino-slepton loops, and thus the impact of ${\rm Arg}(\mu M_1)$ and ${\rm Arg}(\mu A_e)$ on electron EDM becomes unimportant. So we set these two phases to $0$ and focus on the impact of the dominant phase ${\rm Arg}(\mu M_2)$. The tri-linear $A$ parameters for the smuon and selectron are usually assumed to be very small in magnitude, which are set to 0 in our scan. The lower bound 100 GeV in the first line and the lower bound $2 <\tan\beta$ in the third line 
   mainly come from the LEP2 constraints.
   
   For the situation where $\text{Arg}(\mu M_1)$ and $\text{Arg}(\mu A_e)$ dominate,
   we give $|\mu|$ a relatively large value so that the left-right-slepton-bino loop becomes the main contribution of $\delta a_\mu$ and $d_e$. Again we set the unimportant phase ${\rm Arg}(\mu M_2)$ to 0.  However, $|A_e|$ can be as small as 0, and thus $\text{Arg}(\mu A_e)$ will be completely unrestricted. In order to highlight the key information about $\text{Arg}(\mu A_e)$, we will consider two cases with $A_e/\exp(i{\rm Arg}(A_e))=\pm 500~{\rm GeV}$ and $A_e/\exp(i{\rm Arg}(A_e))=\pm 1000~{\rm GeV}$ (note that $A_e/\exp(i{\rm Arg}(A_e))$ can be negative because we take the CP phase ${\rm Arg}(A_e)$ in $(-\pi/2,\pi/2)$). The parameter region is
   \begin{align*}
       &100~\text{GeV}<(|M_1|,|M_2|,M_{E1}=M_{E2},M_{L1}=M_{L2})<2~\text{TeV},\\
       &5~\text{TeV}<|\mu|<10~\text{TeV},\\
       &2<\tan\beta <50,\\
       &\text{Arg}(\mu M_2)=0,\\
       &-\frac{\pi}{2} < \text{Arg}(\mu M_1),\text{Arg}(\mu A_e)<\frac{\pi}{2},\\
       &A_e/\exp(i{\rm Arg}(A_e))=\pm 500~{\rm GeV}\quad \text{or}\quad \pm1000~{\rm GeV}.
   \end{align*}
   
   \begin{figure}[ht]
      \centering 
      \includegraphics[width=.49\textwidth]{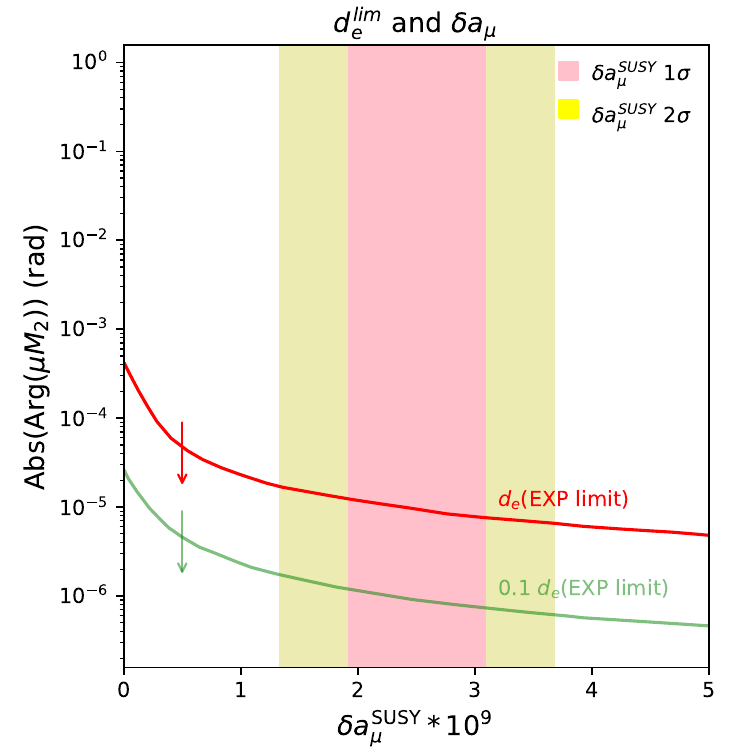}
      \hfill
      \includegraphics[width=.49\textwidth]{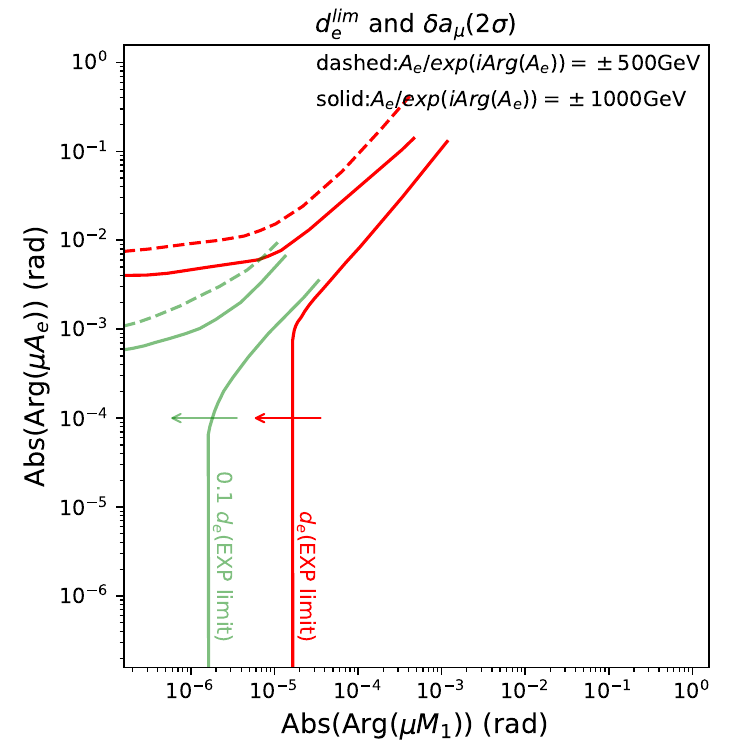}
      \caption{\label{fig:universality} 
      The exclusion curves obtained from the interplay of the electron EDM limit and the muon $g-2$, assuming the lepton flavor universality. In the left panel the region below each curve is the allowed region, while in the right panel the region surrounded by the two branches of each curve is the allowed region. The curves labelled by $d_e$(EXP limit) are obtained by using the current electron EDM experimental limit in Eq.\eqref{eq:electronEDM}. The $2\sigma$ constraint on $\delta a_\mu$ is considered in the right panel. }
   \end{figure}
   
   The scan results are shown in Fig.~\ref{fig:universality}. The phase upper limit curves shown in the figures are obtained by projecting the high-dimensional exclusion surface to two-dimensional planes, which represent the most conservative exclusion limits. In the right panel we considered the $2\sigma$ constraint of $\delta a_\mu$. When $\text{Arg}(\mu M_2)$ dominates, the upper limit of $\text{Arg}(\mu M_2)$ is $1.8\times 10^{-5}(\text{rad})$ for $\delta a_\mu$ within $2\sigma$ range and $1.3\times 10^ {-5}(\text{rad})$ for $\delta a_\mu$ within $1\sigma$ range. If $d^{\rm lim}_e$ is lowered by one order in the future, the upper limit of the corresponding phase is also lowered by an order. From the right panel we see that there exist an accidental cancellation for the CP-phase effects. When this accidental cancellation occurs, even if the phases are rather large, they can still provide a sufficiently small EDM for the electron. Of course, 
   such an accidental cancellation needs a fine-tuning of the relevant parameters, which may indicate some unknown symmetry.  
  Without accidental cancellations, the upper limit of $\text{Arg}(\mu M_1)$ is $1.9\times 10^{-5}$ while 
  the upper limit of $\text{Arg}(\mu A_e)$ is $1.1\times 10^{-2}$ for $A_e/\exp(i{\rm Arg}(A_e))=\pm 500~{ \rm GeV}$ and $5.0\times 10^{-3}$ for $A_e/\exp(i{\rm Arg}(A_e))=\pm 1000~{\rm GeV}$. We see that limit on $\text{Arg}(\mu A_e)$ is two to three orders weaker than $\text{Arg}(\mu M_1)$. 

   \subsection{CP-phase upper limits without flavor universality}
   Now we turn to the case of lepton flavor non-universality and calculate the phase limits from the electron $g-2$ result in Eq.\eqref{eq:electronGM2} (obtained from the Berkeley measurement of $\alpha_{\rm em}$) and the electron EDM limit. We use Eq.\eqref{eq:phaselimit} again to estimate the limit on the dominated phase
   \begin{equation}\label{eq:piliephaselimit}
       |\varphi|\lesssim\frac{4m_e}{|\delta a_e|} \frac{d_e^{\rm lim}}{\rm e}\sim 1\times10^{-6}(\text{rad}),
   \end{equation}
    where we take the central value in Eq.\eqref{eq:electronGM2} for $\delta a_e$ and  the upper bound in Eq.\eqref{eq:electronEDM} for $d_e^{\rm lim}$. 
    
    In this case, the parameter space in the MSSM that can explain both electron $g-2$ and muon $g-2$ is proved to be very narrow~\cite{Badziak:2019gaf, Li:2021koa}. Based on the results in \cite{Badziak:2019gaf, Li:2021koa}, we only have two parameter scenarios to consider (we set the unimportant phases to $0$). The first scenario is~\cite{Badziak:2019gaf}
   \begin{align*}
       & 100~\text{GeV}<(|M_2|,M_{L1},M_{E1},M_{L2})<500~\text{GeV}, \\
       &-300~\text{GeV}<M_1/\exp(i{\rm Arg}(M_1))<-80~\text{GeV},\\
       & 500~\text{GeV}<|\mu|<5~\text{TeV},\\
       &1~\text{TeV}<M_{E2}<5~\text{TeV},\quad 2<\tan\beta <50,\\
       &\text{Arg}(\mu M_2)=0,\\
       &-\frac{\pi}{2}<\text{Arg}(\mu M_1),\text{Arg}(\mu A_e)<\frac{\pi}{2},\\
       &A_e/\exp(i{\rm Arg}(A_e))=\pm 100~{\rm GeV}\quad \text{or}\quad \pm500~{\rm GeV},
   \end{align*}
   In this scenario the electron EDM mainly depends on $\text{Arg}(\mu M_1)$ and $\text{Arg}(\mu A_e)$. The second scenario is~\cite{Li:2021koa}
   \begin{align*}
       & 100~\text{GeV}<(|\mu|,M_{L1},M_{E2})<500~\text{GeV}, \\
       &-200~\text{GeV}<M_1/\exp(i{\rm Arg}(M_1))<-80~\text{GeV},\\ 
       & -2~\text{TeV}<M_2/\exp(i{\rm Arg}(M_2))<-100~\text{GeV},\\
       &M_{E1}=M_{L2}=5~\text{TeV},\quad 2<\tan\beta <50,\\
       &-\frac{\pi}{2}<\text{Arg}(\mu M_1),\text{Arg}(\mu M_2)<\frac{\pi}{2},\\
       &A\text{ parameters} = 0,
   \end{align*}
    In this scenario the electron EDM mainly depends on $\text{Arg}(\mu M_1)$ and $\text{Arg}(\mu M_2)$.
   
   \begin{figure}[ht]
      \centering 
      \includegraphics[width=.49\textwidth]{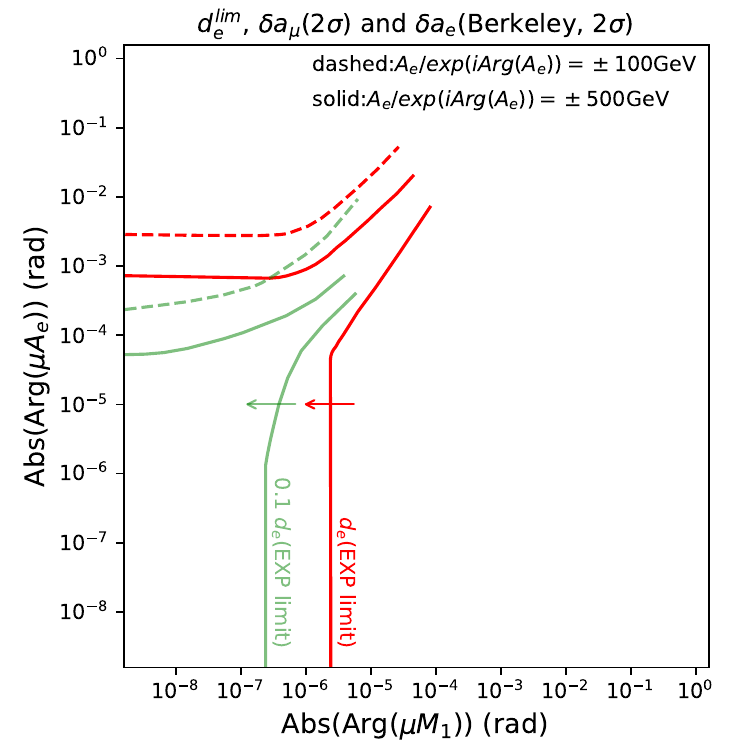}
      \hfill
      \includegraphics[width=.49\textwidth]{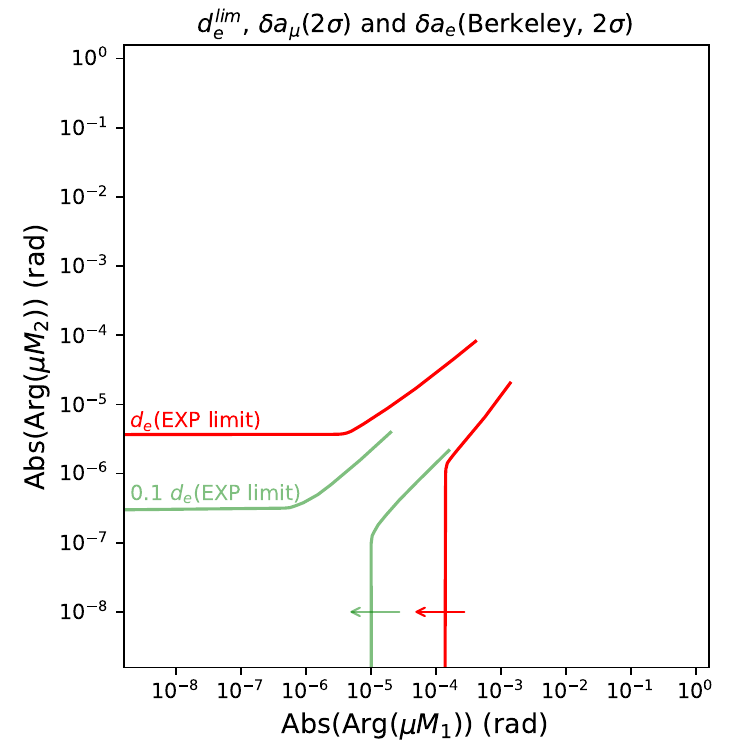}
      \caption{\label{fig:non-universality} 
      The exclusion curves obtained from the electron EDM limit combined with $\delta a_{\mu}$ and $\delta a_e$ (Berkeley). The region surrounded by the two branches of each curve is the corresponding allowed region. The curves labelled by $d_e$(EXP limit) is obtained by using the current electron EDM experimental limit in Eq.\eqref{eq:electronEDM}. }
   \end{figure}
   
   The results of the parameter scan are shown in Fig.~\ref{fig:non-universality}, where the $2\sigma$ constraints of $\delta a_e$ (Berkeley) and $\delta a_\mu$ combined with the electron EDM limit are considered. The accidental cancellation occurs in both cases. Without such accidental cancellation, in the first scenario $\text{Arg}(\mu M_1)$ is restricted to be smaller than $2.4\times 10^{-6}(\text{rad})$ regardless of the value of $|A_e|$, while the upper limit of $\text{Arg}(\mu A_e)$ is $4.1\times 10^{-3}(\text{rad})$ for $A_e/\exp(i{\rm Arg}(A_e))=\pm 100{\rm GeV}$ and $1.0\times 10^{-3}(\text{rad})$ for $A_e/\exp(i{\rm Arg}(A_e))=\pm 500{\rm GeV}$; in the second scenario, the upper limit of $\text{Arg}(\mu M_2)$ is $3.9\times 10^{-6}(\text{rad})$ and the upper limit of  $\text{Arg}(\mu M_1)$ is $1.3\times 10^{-4}(\text{rad})$. 
   In the second scenario, $\delta a_e$ mainly comes from the wino-higgsino-sneutrino loops. Thus $\text{Arg}(\mu M_2)$ dominates $d_e$, and its upper limit is close to the estimated value in  Eq.\eqref{eq:piliephaselimit}. 
   
   Finally, we make some comments. 
   As mentioned in the introduction, we concentrated on the CP-phase effects in the MSSM. 
   We did not study the dark matter issue. In the allowed parameter space displayed in our results, if the dark matter is assumed to be the lightest neutralino, then the dark matter relic density and the direct detection limits may not be easily satisfied. However, the dark matter problem can be avoided if we assume a superWIMP to be the dark matter candidate, which is produced from the late decay of the thermal freeze-out of the lightest neutralino ~\cite{Feng:2003xh}. Since the superWIMP can be much lighter than the lightest neutralino, the relic density upper bound can be readily satisfied. Of course, the superWIMP interacts with nucleon superweakly and thus the direct detection limits can be easily satisfied. Such a superWIMP can be predicted in the framework of SUSY, e.g., a pseudo-goldstino in multi-sector SUSY breaking scenario ~\cite{Argurio:2011hs,Dai:2021eah} with some peculiar phenomenology at the LHC \cite{Liu:2014lda,Hikasa:2014yra}. 
     Note that these CP-phases in the MSSM, already stringently constrained to be very small in magnitude, may not have sizable effects at the colliders like the LHC. However, as shown in this study and in the literature \cite{Athron:2021iuf,Abdughani:2019wai,Li:2021koa,Cox:2018qyi,Badziak:2019gaf}, the interpretation of $(g-2)_{\mu}$ or both $(g-2)_e$ and $(g-2)_\mu$ in the MSSM can severely restrain the parameter space, which requires relatively light electroweakinos and sleptons. Such a restrained parameter space is found to be accessible at the LHC \cite{Athron:2021iuf,Abdughani:2019wai,Li:2021koa,Cox:2018qyi,Badziak:2019gaf}, which satisfies the current LHC constraints  and can be further explored at the HL-LHC \cite{Abdughani:2019wai} or a 100 TeV hadron collider \cite{Kobakhidze:2016mdx}. In addition, the required light electroweakinos may alter the Higgs couplings by a couple of 
   percent which may be tested at a Higgs factory \cite{Abdughani:2019wai}.
   For similar studies in other non-SUSY models or extensions of the MSSM, see, e.g., \cite{Sabatta:2019nfg,Zhu:2021vlz,Abdughani:2021pdc}. 
   Another point we want to stress is that we worked in the phenomenological MSSM and did not discuss the fancy GUT-constrained SUSY models like mSUGRA which are usually hard to explain the muon $g-2$ anomaly and some extensions are needed \cite{Wang:2021bcx,Li:2021pnt,Akula:2013ioa,Wang:2015rli,Wang:2018vrr,Han:2020exx}. 
   
   \section{Conclusions}\label{sec:conclusions}
   In this work, we studied the upper limits of the CP-phases from the constraints of muon $g-2$ and electron $g-2$ combined with the electron EDM experimental limit in the MSSM. We considered two cases:
   one is that assuming lepton flavor universality, we used the muon $g-2$ combined with the electron EDM experimental limit to set constraints on the CP-phases; the other is without assuming lepton flavor universality, we used the electron $g-2$ (Berkeley) and the muon $g-2$ combined with the electron EDM experimental limit to set constraints on the CP-phases. 
   We first derived an approximate analytical expression for the CP-phase upper limits.
   Then we performed a scan over the parameter space to obtain more accurate upper limits.
   We found that with the universality of lepton flavor the muon $g-2$ (FNAL+BNL $2\sigma$) combined with  the electron EDM experimental limit can typically constrain the CP-phases 
   under $1.9\times 10^{-5} (\text{rad})$, while without the lepton flavor universality the electron $g-2$ (Berkeley $2\sigma$) and the muon $g-2$ (FNAL+BNL $2\sigma$) combined with the electron EDM experimental limit can typically constrain the CP-phases under $3.9\times 10^{-6}(\text{rad})$,
   except that some accidental cancellation may relax the limits by a couple of orders.
   Such strong limits on the CP-phases in the MSSM may have some indications for the model building of SUSY.

\addcontentsline{toc}{section}{Acknowledgments}
\acknowledgments
We thank Chengcheng Han and Peiwen Wu for helpful discussions. This work was supported by the National
Natural Science Foundation of China (NNSFC) under grant Nos. 11821505 and 12075300,
by Peng-Huan-Wu Theoretical Physics Innovation Center (12047503), by the CAS Center
for Excellence in Particle Physics (CCEPP), by the CAS Key Research Program of Frontier
Sciences, and by a Key R\&D Program of Ministry of Science and Technology of China under
number 2017YFA0402204, and by the Key Research Program of the Chinese Academy of
Sciences, grant No. XDPB15.

\addcontentsline{toc}{section}{References}
\bibliographystyle{JHEP}
\bibliography{bibliography}

\end{document}